\documentclass[preprint,amsmath,amssymb,aps]{revtex4-1}

\usepackage{graphicx}
\usepackage{dcolumn}
\usepackage{bm}
\usepackage{braket}

\usepackage{amsmath}	
\begin{document}

\title{Real-time evolution method and its application to 3$\bm \alpha$ cluster system }
\author{R. Imai}
 \affiliation{Department of Physics, Hokkaido University, 060-0810 Sapporo, Japan}
\author{M. Kimura }
 \email{masaaki@nucl.sci.hokudai.ac.jp}
 \affiliation{Department of Physics, Hokkaido University, 060-0810 Sapporo, Japan}
 \affiliation{Nuclear Reaction Data Centre, Faculty of Science, Hokkaido University,
  060-0810 Sapporo, Japan} 
\author{T. Tada}
 \affiliation{Department of Physics, Hokkaido University, 060-0810 Sapporo, Japan}

\date{\today}

\begin{abstract}
 A new theoretical method is proposed  to  describe the ground and excited cluster states of
 atomic nuclei.  The method utilizes the equation-of-motion of the Gaussian wave packets to
 generate the basis wave functions having  various  cluster configurations. The generated basis
 wave functions are superposed to diagonalize  the Hamiltonian. In other words, this method uses
 the real time as the generator coordinate.   The application to the $3\alpha$ system as a
 benchmark shows  that the  new method works efficiently and yields the result consistent
 with or better than  the other cluster models. Brief discussion on the structure of the
 excited $0^+$ and $1^-$  states is also made.  
\end{abstract}

\pacs{Valid PACS appear here}
\maketitle


\section{Introduction}
It has long been known that the Hoyle state (the $0^+_2$ state of $^{12}{\rm C}$) is a dilute
gas-like $\alpha$ cluster state dominated by the $s$-wave \cite{Uegaki1978,Fujiwara1980,
Kamimura1981,Descouvemont1987,Kanada-Enyo1998,Descouvemont2003,Chernykh2007,Kanada-Enyo2007a}.
Later, it was pointed out that the  Hoyle state can be regarded as a Bose-Einstein condensate of
$\alpha$ particles \cite{Tohsaki2001,Yamada2012,Schuck2016}. This findings motivated many studies
on $^{12}{\rm C}$ and the related topics. For example, the idea of the $\alpha$ particle
condensate has been extended to other excited states above the Hoyle state. Namely, the $2^+$
state at 10.03 MeV \cite{Freer2009,Itoh2011,Zimmerman2011,Zimmerman2013}  and  the $4^+$ state
at 13.3 MeV \cite{Freer2011} are considered as the members of the "Hoyle band"
\cite{Freer2014a,Funaki2015,Schuck2016}. More recently, the $0^+_3$ state at 10.3 MeV 
\cite{Itoh2011,Itoh2013a} is suggested as  the "breathing mode" of the Hoyle state
\cite{Kurokawa2005, Kurokawa2007, Ohtsubo2013,Funaki2015,Funaki2016,Zhou2016}. The  possible
formation of the 3$\alpha$ linear-chain ($0^+_4$ state) has also been discussed 
\cite{Funaki2015,Funaki2016}. 

The discussion  has been naturally extended to the condensate of many $\alpha$ particles. The
candidates of the 4$\alpha$ condensate in $^{16}{\rm O}$ are under the intensive discussions
\cite{Wakasa2007,Funaki2008,Curtis2013,Rodrigues2014,Bijker2014,Ogloblin2016,Bijker2017,Li2017a}.
However, the theoretical and experimental information for the 5$\alpha$ and more
$\alpha$ particle condensate \cite{Yamada2004,Kokalova2006,Itagaki2007,Schuck2016} is rather
scarce. The clustering of non-$\alpha$ nuclei is another direction of the extension. For example,
the Hoyle-analog states with a nucleon hole or particle are discussed for $^{11}{\rm B}$
\cite{Descouvemont1995,Kanada-Enyo2007,Kawabata2007,Yamada2010} and $^{13}{\rm C}$
\cite{Yamada2006,Yamada2015}. The 3$\alpha$ linear-chains accompanied by the valence neutrons are
expected in neutron-rich C isotopes \cite{Itagaki2001,Suhara2010,Freer2014,Baba2014,Ebran2014,
DellAquila2016,Fritsch2016,Tian2016,Baba2016,Li2017,Yamaguchi2017,Baba2017}. Thus, nowadays the
researches are extending to the highly excited cluster states  composed of many clusters and
nucleons.   

However, when the number of the constituent clusters or nucleons increases, the description of
the cluster states becomes difficult. For example, suppose that one employs the generator
coordinate method (GCM) \cite{Hill1953,Griffin1957} which superposes many basis wave
functions. Then, it is easy to imagine that the number of basis wave functions required for the
description of the cluster states increases very quickly as the number of constituent particles
increases or the system becomes dilute. As a result, much computational power is demanded and the
practical calculation becomes difficult. This may be one of the reason why the 5$\alpha$ and more
$\alpha$ particle condensate are rarely studied based on the microscopic models. Therefore, new
method which efficiently generates the basis wave functions is highly desirable and
indispensable. For this purpose, several methods have been developed such as the stochastic
sampling of the basis  wave functions \cite{Suzuki1998,Itagaki2003,Mitroy2013} and the
imaginary-time development method \cite{Fukuoka2013}.  

In this study, we propose an alternative method which utilizes the equation-of-motion (EOM) of
the Gaussian wave packets. The basis wave functions are generated by the real-time evolution of
the  system governed by the EOM, and they are superposed to diagonalize the Hamiltonian. In other
words, this  method employs the real time as the generator coordinate.  As a benchmark of the 
methodology, we  applied it to the $3\alpha$ system ($^{12}{\rm C}$). It is shown that the  new
method works efficiently and yields the result consistent with or better than  the other
cluster models. Furthermore, based on the isoscalar (IS) monopole and dipole transition strengths, 
we briefly discuss the structure of the excited $0^+$ and $1^-$ states.

This paper is organized as follows. In the next section, we explain the framework of the new
method named real-time evolution method (REM). In the section \ref{sec:result}, we present the
result of the numerical benchmark. We also discuss the structure of the excited $0^+$ and $1^-$
states briefly. The final section summarizes this work.

\section{Theoretical framework}
Here, we explain the framework of the real-time evolution method. For simplicity, we assume its
application to the $\alpha$ cluster wave functions ($4N$ nuclei). However, it is noted that the
method is also applicable to more general cases such as non-$\alpha$ cluster wave functions,
antisymmetrized molecular dynamics (AMD) and fermionic molecular dynamics (FMD) wave functions.  

\subsection{Hamiltonian and GCM wave function}
The Hamiltonian for the $N\alpha$ systems composed of 4$N$ nucleons is given as,
\begin{align}
 \hat{H} = \sum_{i=1}^{4N} \hat{t}_i  + \sum_{i<j}^{4N}\hat{v}_N(r_{ij}) +
 \sum_{i<j}^{4N}\hat{v}_C(r_{ij}) - \hat{t}_{cm}, 
\end{align}
where $\hat{t}_i$ and $\hat{t}_{cm}$ respectively denote the kinetic energies of the nucleons  and
the center-of-mass. The $\hat{v}_{N}$ and $\hat{v}_{C}$ denote the effective nucleon-nucleon
interaction and Coulomb interactions, respectively. The parameter set of $\hat{v}_N$ is explained
later. 

As for the intrinsic wave function of $N\alpha$ system, we employ the Brink-Bloch wave function
\cite{Brink1966} which is composed of $\alpha$ clusters having $(0s)^4$ configurations,
\begin{align}
 &\Phi(\bm Z_1,...,\bm Z_N) = \mathcal A
 \Set{\Phi_\alpha(\bm Z_1)\cdots\Phi_\alpha(\bm Z_N)},\label{eq:brink1}\\
 &\Phi_\alpha(\bm Z) = \mathcal A
 \Set{\phi(\bm r_1,\bm Z)\chi_{p\uparrow}\cdots\phi(\bm r_4,\bm Z)\chi_{n\downarrow}},\\
 &\phi(\bm r,\bm Z) = \Bigl(\frac{2\nu}{\pi}\Bigr)^{3/4}\exp
 \Set{-\nu\Bigl(\bm r- \frac{\bm Z}{\sqrt{\nu}}\Bigr)^2+\frac{1}{2}Z^2}.
\end{align}
Here $\Phi_\alpha(\bm Z)$ denotes the wave packet describing the $\alpha$ cluster located at
$\bm Z$. The three-dimensional vectors $\bm Z_1,...,\bm Z_N$ are complex numbered and describe
the $\alpha$ cluster positions in the phase space. 

Similarly to other cluster models, we superpose the intrinsic wave function having different
configurations (different sets of the complex vectors $\bm Z_1,...,\bm Z_N$) after the parity and
angular momentum projection (GCM).  The most general form of the GCM wave function may be written
as, 
\begin{align}
 \Psi^{J\pi}_M=&\sum_{K}\int d^3 Z_1...d^3 Z_N\nonumber\\
 &\times f_K(\bm Z_1,...,\bm Z_N)
 \hat{P}^{J\pi}_{MK}\Phi(\bm Z_1,...,\bm Z_N),
 \label{eq:gcmwf1}
\end{align}
where $\hat{P}^{J\pi}_{MK}$ is the parity and angular momentum projector. The amplitude of the
superposition $f_K(\bm Z_1,...,\bm Z_N)$ must be determined in some ways. 
For example, the original THSR wave function ($J=M=K=0$) \cite{Tohsaki2001}
 asserts that the amplitude can be written as  
\begin{align}
 f_0(\bm R_1,...,\bm R_N) = \prod_{i=1}^N \exp\set{-\frac{1}{2\beta^2}R_i^2},
\end{align}
where the vectors $\bm Z_1,...,\bm Z_N$ are reduced to the real valued vectors 
$\bm R_1,...,\bm R_N$, and the parameter $\beta$ controls the size of the $\alpha$ particle
condensate. It is known that this THSR ansatz works surprisingly well for the ground and
excited $0^+$ states of $^{12}{\rm C}$ \cite{Tohsaki2001,Funaki2003,Yamada2012,Schuck2016}. 

In other ordinary cluster models, Eq. (\ref{eq:gcmwf1}) is often discretized and approximated by a
sum of the finite number of the basis wave functions, 
\begin{align}
 \Psi^{J\pi}_{M}=\sum_{p=1}^{p_{max}}\sum_{K=-J}^J f_{pK}\hat{P}^{J\pi}_{MK}
 \Phi(\bm Z_1^{(p)},...,\bm Z_N^{(p)}),
 \label{eq:gcmwf2}
\end{align}
and the amplitude $f_{pK}$ is calculated by the Griffin-Hill-Wheeler equation 
\cite{Hill1953,Griffin1957}. Here, $\bm Z_1^{(p)},...,\bm Z_N^{(p)}$ denotes the $p$th set of the
vectors  $\bm Z_1,...,\bm Z_N$ and the number of the superposed basis wave function  is equal to
$p_{max}$. If $p_{max}$ is sufficiently large and the set of the vectors
 $\bm Z_1^{(p)},...,\bm Z_N^{(p)}$
covers various configurations of $\alpha$ clusters, Eq. (\ref{eq:gcmwf2}) will be a good
approximation, but the increase of $p_{max}$ requires much computational cost. It is easy to
imagine that the number of basis  wave function $p_{max}$ required for a reasonable description
of $N\alpha$ systems will be greatly increased, when the number of $\alpha$ particle is
increased. This is one of the reason, for example, why the condensation of 5 and more $\alpha$
particles are rarely studied by the microscopic models. 

Therefore, if one employs the approximation given by Eq. (\ref{eq:gcmwf2}), it is
essentially important to find a method which  {\it efficiently} generates the set of the vectors  
$\bm Z_1^{(p)},...,\bm Z_N^{(p)}$  to reduce the computational cost. For this purpose, several
methods such as the stochastic method \cite{Suzuki1998,Itagaki2003,Mitroy2013}  and the imaginary
time evolution methods \cite{Fukuoka2013} have  been proposed, and in this study, we introduce a
new method which uses the real-time evolution of the $\alpha$ particle wave packets.

\subsection{Real-time evolution method}
In the present study, the EOM of the $\alpha$ particle wave
packets is used to generate the sets of the vectors $\bm Z_1^{(p)},...,\bm Z_N^{(p)}$.  By
applying the time-dependent variational principle to the intrinsic wave function given by
Eq. (\ref{eq:brink1}),
\begin{align}
 \delta\int dt\frac{\braket{\Phi(\bm Z_1,...,\bm Z_N)|i\hbar\ d/dt - \hat{H}|
 \Phi(\bm Z_1,...,\bm Z_N)}}
 {\braket{\Phi(\bm Z_1,...,\bm Z_N)|\Phi(\bm Z_1,...,\bm Z_N)}}=0,
\end{align}
one obtains the EOM for the $\alpha$ particle centroids $\bm Z_1,...,\bm Z_N$,
\begin{align}
  &i\hbar \sum_{j=1}^N\sum_{\sigma=x,y,z} C_{i\rho j\sigma}\frac{dZ_{j\sigma}}{dt} = 
 \frac{\partial \mathcal H_{int}}{\partial Z_{i\rho}^*}, \label{eq:eom}\\
 &\mathcal H_{int} \equiv \frac{\braket{\Phi(\bm Z_1,...,\bm Z_N)|\hat{H}
 |\Phi(\bm Z_1,...,\bm Z_N)}}
 {\braket{\Phi(\bm Z_1,...,\bm Z_N)|\Phi(\bm Z_1,...,\bm Z_N)}}, \\
 &C_{i\rho j\sigma} \equiv 
 \frac{\partial^2 \ln\braket{\Phi(\bm Z_1,...,\bm Z_N)|\Phi(\bm Z_1,...,\bm Z_N)}}
 {\partial Z_{i\rho}^*\partial Z_{j\sigma}}.
\end{align}

Starting from an arbitrary initial wave function at $t=0$, we solve the time evolution of $\bm
Z_1,...,\bm Z_N$. As a result, the EOS yields the set 
of the vectors $\bm Z_1(t),...,\bm Z_N(t)$ as function of the time $t$, which defines the wave
function $\Phi(\bm Z_1(t),...,\bm Z_N(t))$ at each time.
Despite of its classical form, this EOM still holds the information of the quantum system.
For example, it was shown that the nuclear phase shift of the $\alpha$-$\alpha$ scattering
can be obtained from the classical trajectory of the wave packet centroids \cite{Saraceno1983}. 
In addition to this, it was shown that the ensemble of the wave functions 
$\Phi(\bm Z_1(t),...,\bm Z_N(t))$ possesses following good properties
\cite{Schnack1996,Ono1996,Ono1996a}.
\begin{enumerate}
 \item The ensemble of the time-dependent wave functions 
       $\Phi(\bm Z_1(t),...,\bm Z_N(t))$ has ergodic nature
 \item And it follows the {\it quantum} statistics
\end{enumerate}
if the nucleon-nucleon collisions and nucleon emission processes are properly treated.
Indeed, on the basis of this EOM properties, the nuclear liquid-gas phase transition during the
heavy-ion collisions and the caloric curve for the finite nuclei have been studied
\cite{Schnack1997,Sugawa1999,Furuta2006,Kanada-Enyo2012}.
Therefore, we  expect that, if time is evolved long enough, the ensemble of the wave functions 
$\Phi(\bm Z_1(t),...,\bm Z_N(t))$  spans a good model space for $N\alpha$ systems. In other words,
we expect that the bound and resonant states of $N\alpha$ systems are reasonably described by the 
superposition of the basis wave functions as follows,
\begin{align}
 \Psi^{J\pi}_{M}(T)=\int_0^{T} dt\sum_{K=-J}^J\hat{P}^{J\pi}_{MK}\biggl\{
 f_{K}(t) \Phi(\bm Z_1(t),...,\bm Z_N(t))\nonumber\\
 + g_{K}(t)\Phi(\bm Z_1^*(t),...,\bm Z_N^*(t))\biggr\}. \label{eq:gcmwf3}
\end{align}
Here, the complex conjugated basis wave functions are also superposed to properly describe
time-even states. The coefficients $f_{K}(t)$ and $g_K(t)$ should be determined by the
diagonalization of the Hamiltonian. Eq. (\ref{eq:gcmwf3}) can be regarded as the GCM wave function
which employs the real-time $t$ as the generator coordinate.

\subsection{Numerical calculation}
In this study, REM calculation is performed for $3\alpha$ cluster system ($^{12}{\rm C}$). 
For the sake of the comparison, we used the Volkov No. 2 effective nucleon-nucleon interaction
\cite{Volkov1965} with a slight modification \cite{Fujiwara1980,Kamimura1981}, which is common to
the other studies using resonating group method (RGM) \cite{Fujiwara1980,Kamimura1981} and
Tohsaki-Horiuchi-Schuck-R\"opke (THSR) wave function \cite{Funaki2003,Funaki2015,Funaki2016}.  The
numerical calculation was performed in the following steps. 

(1) In the first step, we randomly
generate $3\alpha$ cluster wave function and calculate the {\it imaginary-time} evolution of the
system, 
\begin{align}
 &i\hbar\frac{d\bm Z_{i}}{d\tau} = \mu
 \frac{\partial \mathcal H_{int}}{\partial \bm Z_{i}^*}, \label{eq:ieom}
\end{align}
where $\mu$ is an arbitrary negative number. Eq. (\ref{eq:ieom}) decreases the  intrinsic energy
$\mathcal H_{int}$, as the imaginary time $\tau$ is evolved. The imaginary-time evolution is
continued until the intrinsic excitation energy,
\begin{align}
 E^*_{int} = \mathcal H_{int} - \mathcal H_{int}^{min},
\end{align}
equals to a certain value. Here, $\mathcal H_{int}^{min}$ is the minimum intrinsic energy 
obtained by the very long imaginary-time evolution, which is -74.5 MeV in the present Hamiltonian. 
In the practical calculation, we tested several values of $E^*_{int}$ (10, 20, 25 and 30 MeV) and
found that $E^*_{int}= 25$ MeV results in the best convergence.


(2) In the second step, we calculate the {\it real-time} evolution (Eq. (\ref{eq:eom})) starting from
the initial wave function obtained in the first step. For the numerical calculation, the time is
discretized with an interval of $\Delta t=0.02$ fm/c, 
\begin{align}
 t_p = (p-1) \Delta t, \quad p=1,2,..., p_{max},
\end{align}
and the maximal propagation time is $T_{max} = (p_{max}-1) \Delta t=6,000 $ fm/c. It is noted that
the intrinsic energy $\mathcal H_{int}$, and hence $E^*_{int}$, is conserved by the EOM.  As a
result, the time evolution calculation yields a set of the Brink-Bloch wave functions 
$\Phi(\bm Z_1(t_p),...,\bm Z_N(t_p)), \quad p=1,...,p_{max}$ having the same $E^*_{int}$. And it
is used as the basis wave function of the GCM calculation in the next step.

\begin{figure}[h!]
 \begin{center}
  \includegraphics[width=0.8\hsize]{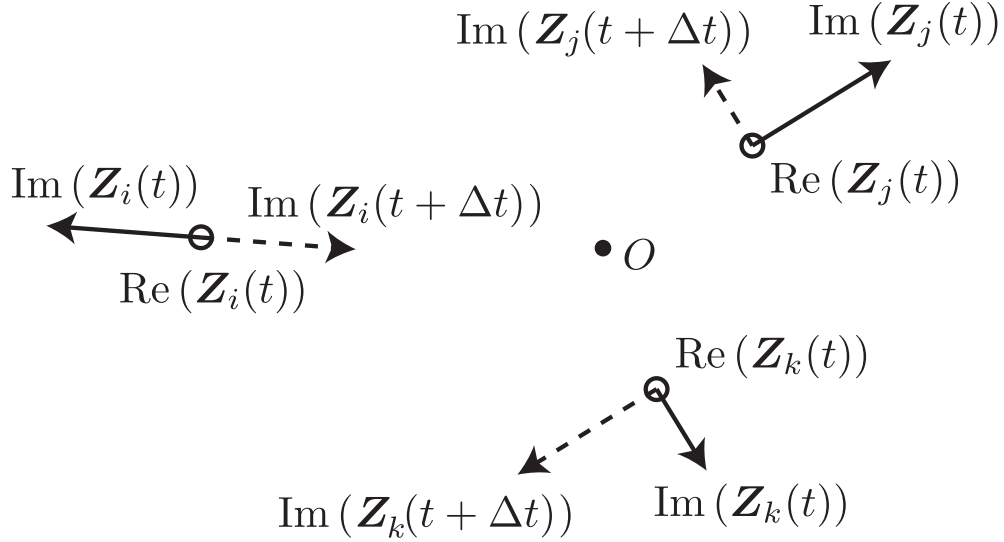}
  \caption{The rebound of $\alpha$ clusters described by
  Eq. (\ref{eq:bounce1})-(\ref{eq:bounce3}). Open circles represent the real-part of $\bm Z$, 
  and the solid (dotted) arrows represent the imaginary part of $\bm Z$ before (after) the 
  rebound. The rebound does not   change the real-parts of $\bm Z_i$, $\bm Z_j$ and $\bm Z_k$
  (positions of  $\alpha$ clusters),   but changes the imaginary-parts (momenta). It reverts the
  momentum of   $\bm Z_i$ and the center-of-mass momentum between $\bm Z_j$ and $\bm Z_k$, but
  conserves the relative   momentum   between $\bm Z_j$ and $\bm Z_k$.}   
  \label{fig:illust}
 \end{center}
\end{figure}

If $E^*_{int}$ is large, $\alpha$ clusters occasionally escape out to infinite distance during the
time evolution. This yields basis wave  functions having unphysically large radius, which are
useless for the  description of the bound or resonant states. To avoid this problem, we impose
additional condition to the calculation. When the condition,
\begin{align}
 \max_{i}\Re\left(|\bm Z_i(t)|/\sqrt{\nu}\right) > R_{max},
\end{align}
is satisfied, {\it i.e.} if any of $\alpha$ clusters is distant more than $R_{max}$, 
we interchange their momentum by hand as follows, 
\begin{align}
 \bm Z_i(t+\Delta t)&= {\rm Re}\left(\bm Z_i(t)\right) - i {\rm Im}\left(\bm Z_i(t)\right),
 \label{eq:bounce1}\\
 \bm Z_j(t+\Delta t)&= {\rm Re}\left(\bm Z_j(t)\right) - i {\rm Im}\left(\bm Z_k(t)\right),
 \label{eq:bounce2}\\
 \bm Z_k(t+\Delta t)&= {\rm Re}\left(\bm Z_k(t)\right) - i {\rm Im}\left(\bm Z_j(t)\right),
 \label{eq:bounce3}
\end{align}
where we assume that $|\bm Z_i(t)|>R_{max}$. It is noted that the real part of $\bm Z(t)$
corresponds the coordinate of $\alpha$ cluster, while the imaginary part corresponds to the
momentum. As a result, $\alpha$ clusters rebound as illustrated in Fig. \ref{fig:illust}. 
In the present calculation, the maximum distance is chosen as $R_{max}=10.0$ fm.

(3) Thus-obtained basis wave functions are superposed by using the real-time $t$ as a generator
coordinate. In the following we call this step GCM calculation. 
Since the time is discretized, Eq. (\ref{eq:gcmwf3}) should read,
\begin{align}
 \Psi^{J\pi}_M =& \sum_{p=1}^{p_{max}}\sum_{K=-J}^J \hat{P}^{J\pi}_{MK}\Bigl\{
 f_K(t_p)\Phi(t_p) +g_K(t_p)\Phi^*(t_p)\Bigr\}, 
\end{align}
Here, the basis wave function $\Phi(\bm Z_1(t_p),...,\bm Z_N(t_p))$ is abbreviated as $\Phi(t_p)$. 
The set of the coefficient $f_K(t_p)$, $g_K(t_p)$ and the eigen-energy are determined by
solving the Griffin-Hill-Wheeler equation to diagonalize the Hamiltonian.

In the practical calculation, a set of basis wave functions $\Phi(t_p)$ obtained by the time
evolution is severely redundant. This makes it difficult to solve the Griffin-Hill-Wheeler
equation accurately. To avoid this problem, we remove the basis  wave functions which have large
overlap with others. When a basis wave function $\Phi(t_p)$ satisfies the following condition, 
\begin{align}
\max_{t<t_p} \frac{|\braket{\Phi(t)| \Phi(t_p)}|^2}
 {\braket{\Phi(t)| \Phi(t)}\braket{\Phi(t_p)| \Phi(t_p)}} > O_{max}, 
\end{align}
it is removed from the ensemble. Namely, we do not use the basis wave functions which have
the overlap with the past wave functions larger than $O_{max}$. In the present calculation
$O_{max}$ is chosen as 0.75. 

(4) As discussed later, above-explained GCM calculation has problem to describe highly excited
broad resonances, because of contamination of the non-resonant wave functions. To
overcome this problem, we apply the $r^2$-constraint method proposed by Funaki {\it et al.}
\cite{Funaki2006}. Following this method, we first diagonalize the radius operator, 
\begin{align}
 &\sum_{K'q}\braket{\hat{P}^{J\pi}_{MK}\Phi(t_p)|\hat{r}^2-r_a^2|\hat{P}^{J\pi}_{MK'}
 \Phi(t_q)}e_{K'qa} = 0, \\
 &\hat{r}^2 = \sum_{i=1}^{4N}(\bm r_i - \bm r_{cm})^2/(4N),
\end{align}
which defines a new set of the basis wave functions,
\begin{align}
 \widetilde{\Phi}^{J\pi}_{Ma}= \sum_{Kp}e_{Kpa}P^{J\pi}_{MK}\Phi(t_p),
\end{align}
corresponding to the eigenvalue $r_a^2$. Superposing these new basis, we construct
the $r^2$-constrained GCM wave function, 
\begin{align}
 \Psi^{J\pi}_{M}=  \sideset{}{'}\sum_{a (r^2_a < r^2_{cut})} \left\{
\widetilde{f}_a\widetilde{\Phi}^{J\pi}_{Ma} + \widetilde{g}_a\widetilde{\Phi}^{J\pi*}_{Ma}
 \right\}. 
\end{align}
Here $\sum'$ denotes the conditional summation running over all $a$ which satisfy the condition  
$r_a^2 < r^2_{cut}$. Namely, the basis wave functions which have too large eigenvalue of the
radius operator are excluded. The coefficients $\widetilde{f}_a$, $\widetilde{g}_a$ and the 
eigen-energies are determined by solving the Griffin-Hill-Wheeler equation.
It has been shown that this method effectively separates the
resonant states from the non-resonant states. In the present calculation, the cut-off radius
$r^2_{cut}$ is varied ranging from $(5.0\ \rm fm)^2$  to $(7.0\ \rm fm)^2$ to check the
convergence.

\section{Numerical Results}\label{sec:result}
In this section, we explain how our method works, and compare the obtained result with those of
other models to check the validity and efficiency of the REM. The detailed discussion on the
structures of cluster states in $^{12}{\rm C}$ will be made in our forthcoming work.

\subsection{Real-time evolution}
As explained in the previous section, the REM relies on the ergotic nature of the EOM. Therefore,
if the time is propagated long enough, the results should be converged and should not depend on
the initial wave functions. To check these points, we tested two different 
initial wave functions with $E^*_{int}=25$ MeV to yield ensembles of the wave functions, which we
denote set 1 and 2.
\begin{figure}[h!]
 \begin{center}
  \includegraphics[width=\hsize]{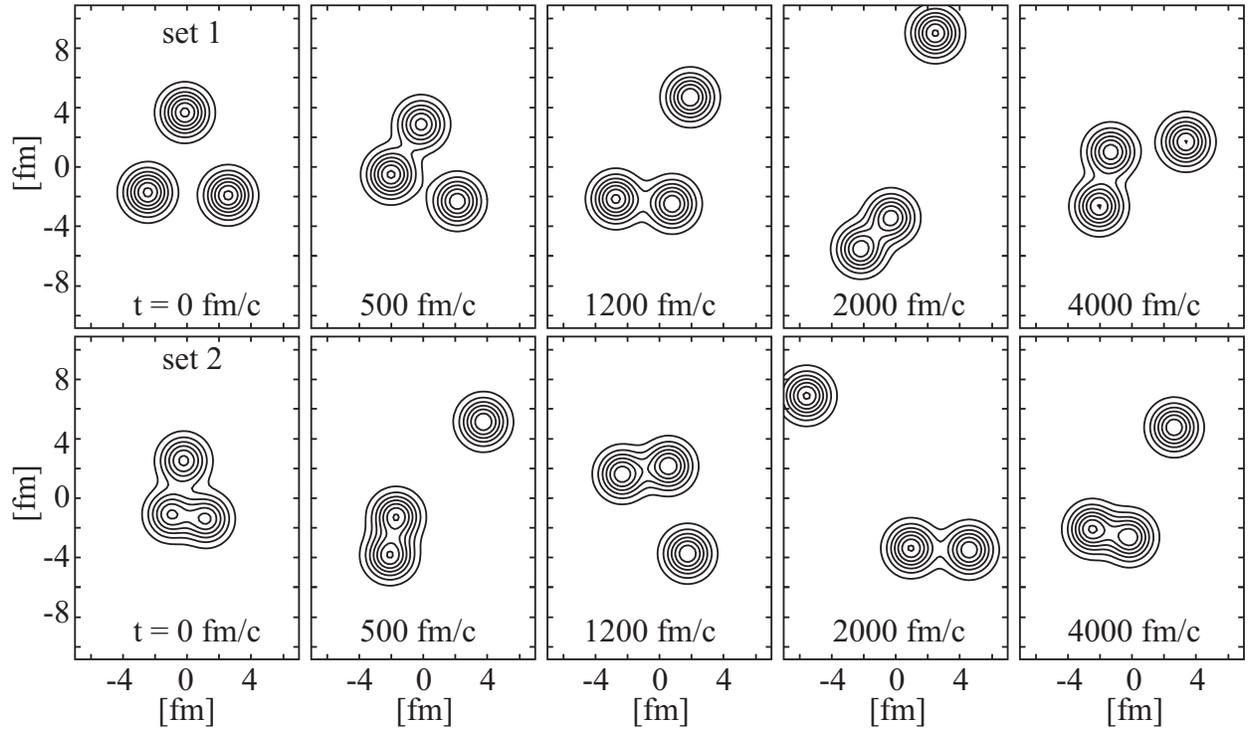}
  \caption{Intrinsic density snapshots at the propagation time $t=0$, 500, 1200, 2000 and
  4000 $\rm fm/c$. Upper (lower) panels show the ensemble of the wave functions obtained
  by the EOM starting from the wave function at $t=0$ $\rm fm/c$.}  
  \label{fig:density}
 \end{center}
\end{figure}

\begin{figure}[t!]
 \begin{center}
  \includegraphics[width=0.9\hsize]{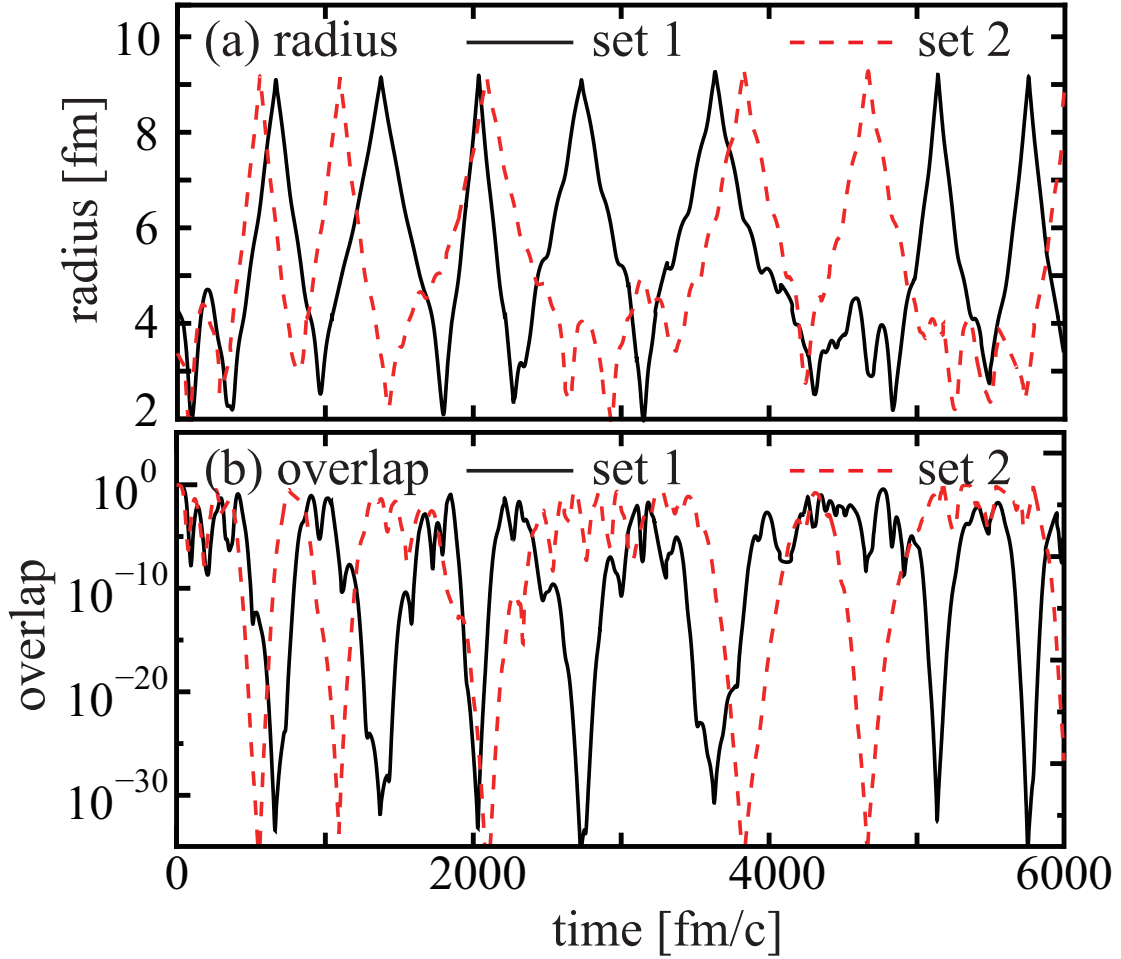}
  \caption{(a) Radius of the intrinsic wave function as function of time. (b) Overlap between the
  wave functions at $t=0$ and $t$ after the projection to the $J^\pi=0^+$.}
  \label{fig:timedev}
 \end{center}
\end{figure}

To illustrate how the 3$\alpha$ system is evolved by the EOM, Fig. \ref{fig:density} shows the
several wave functions of set 1 and 2 at particular times.
In both ensembles, disregard to the initial condition, $\alpha$ clusters distribute in 
various ways; they are close to each other in some time and far distant in another time.
Actually, the system repeats spatial expansion and contraction as time being evolved, which can be
confirmed from the radius of the system as function of time shown in Fig. \ref{fig:timedev} (a). 
It is noted that the unphysical change of the expansion velocity at the maximum radius around 9 fm
is because of the artificial rebound of the $\alpha$ clusters described by
Eqs. (\ref{eq:bounce1})-(\ref{eq:bounce3}).   
Figure \ref{fig:timedev} (b) shows the squared overlap between the wave function $\Phi(t)$ and
initial wave function $\Phi(0)$ after the projection to $J^\pi=0^+$ state, that is defined as,
\begin{align}
 {\mathcal O}(t) = \frac{|\braket{\hat P^{0^+}\Phi(0)|\hat P^{0^+}\Phi(t)}|^2}
 {\braket{\hat P^{0^+}\Phi(0)|\hat P^{0^+}\Phi(0)}
 \braket{\hat P^{0^+}\Phi(t)|\hat P^{0^+}\Phi(t)}}. 
\end{align}
We see that the overlap is rather small, and hence, the wave function is almost orthogonal to the
initial wave function in most of the time. Thus,  the EOM generates various $\alpha$ cluster
configurations automatically.

\subsection{Convergence without and with $\bm {r^2}$-constraint}
\begin{figure}[h]
 \begin{center}
  \includegraphics[width=0.9\hsize]{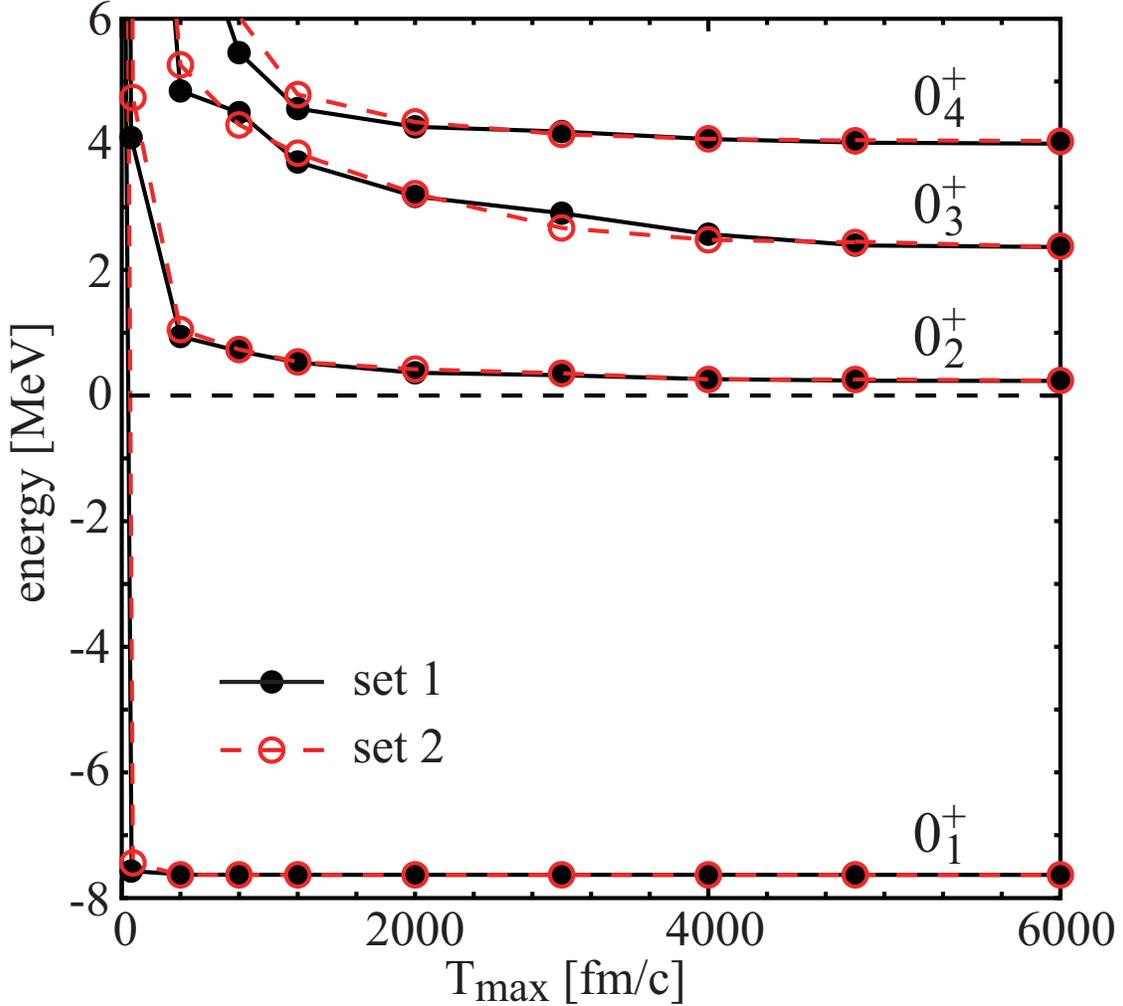}
  \caption{(Color online) The eigen-energies of the $0^+$ states measured from the $3\alpha$
  threshold as function of the propagation time $T$ obtained by the REM without the
  $r^2$-constraint. Open and filled symbols show the results obtained by using the different
  initial wave functions at $t=0$ fm/c.}  
  \label{fig:energy_nocnst}
 \end{center}
\end{figure}

We first discuss the GCM results obtained without $r^2$-constraint. 
Figure  \ref{fig:energy_nocnst} shows the energies the $0^+$ states as functions of the maximum
propagation time $T_{max}$. 
\begin{figure}[h]
 \begin{center}
  \includegraphics[width=0.9\hsize]{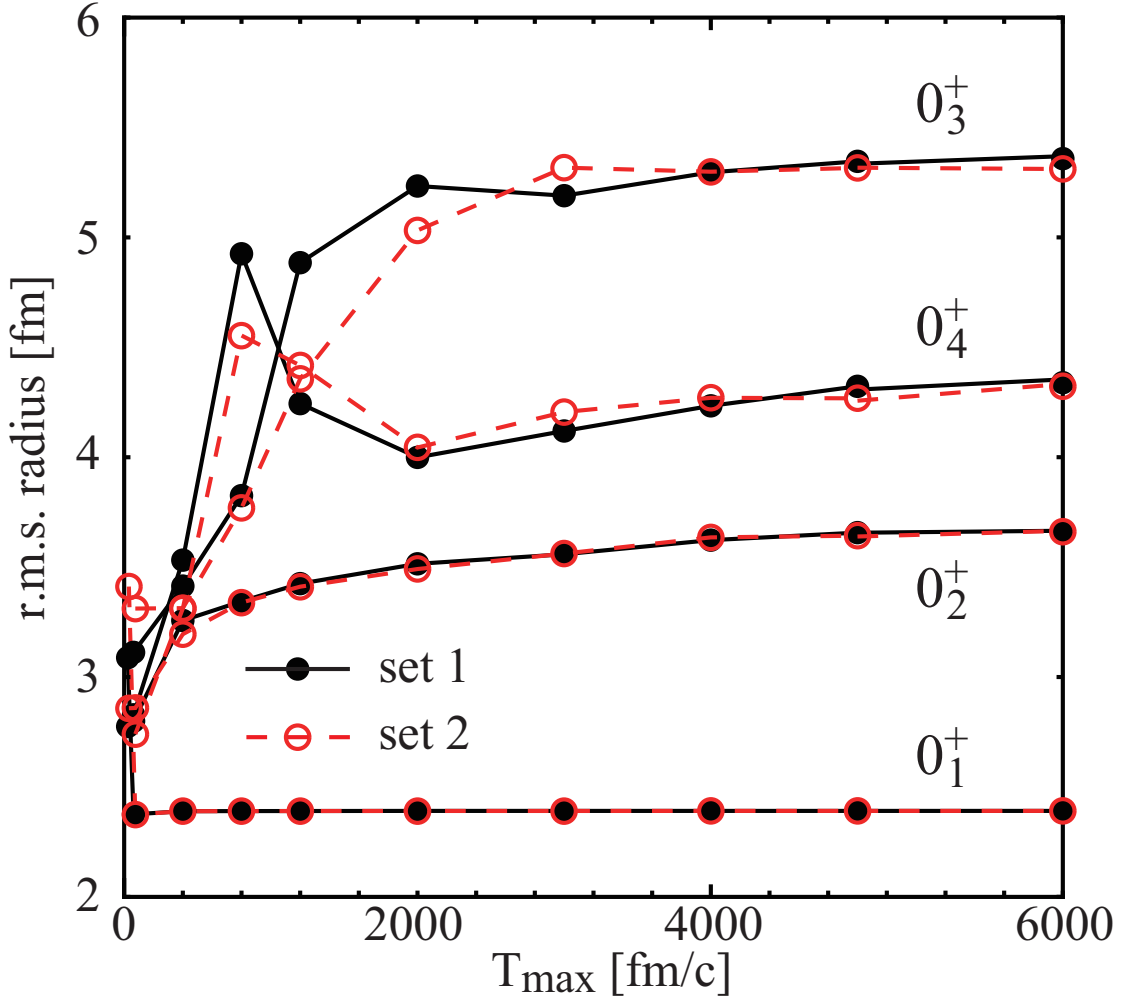}
  \caption{(Color online) Same with Fig. \ref{fig:energy_nocnst} but for radii.}
  \label{fig:radius_nocnst} 
 \end{center}
\end{figure}
We see that the energies of the $0^+$ states converge and are independent of the initial wave
functions, if the propagation time is long enough. In particular, the energy of the ground state
converges very quickly, despite of the rather high intrinsic excitation energy 
($E^*_{int}=25$ MeV) of the basis wave functions generated by the EOM. We also found that the
quick convergence is common to another bound state ($2^+_1$ state).

On the other hand, the energy convergence of the excited $0^+$ states is not as fast as that of
the ground state. In particular, it is interesting to note that the convergence of the $0^+_3$
state looks much slower than others. This is related to the fact that the $0^+_3$ state
is a very broad resonance \cite{Itoh2011,Itoh2013a}. Furthermore, if we observe the figure
carefully, we find that the energies of these unbound states still go down even at large
$T_{max}$. This is because of the contamination of the non-resonant wave functions to these
excited $0^+$ states, which can be seen more clearly in the $T_{max}$ dependence of the radius
shown in Fig. \ref{fig:radius_nocnst}. Again we see that the convergence of the ground state is 
surprisingly fast, but the unbound states are not. In this figure, we clearly observe that the
radii of the unbound states continuously increases showing the contamination of
non-resonant wave functions with huge radii. 

To avoid the contamination of the non-resonant wave functions, we applied the $r^2$-constraint
\cite{Funaki2006}.  This prescription excludes the basis wave function with huge
radius and makes it possible to obtain approximate energies and wave functions of the resonant
states. Since the $r^2$-constraint was already applied to THSR wave function
\cite{Funaki2015,Funaki2015,Zhou2016}, it is worthwhile to compare the results between THSR and
REM with $r^2$-constraint.  
\begin{figure}[h]
 \begin{center}
  \includegraphics[width=0.85\hsize]{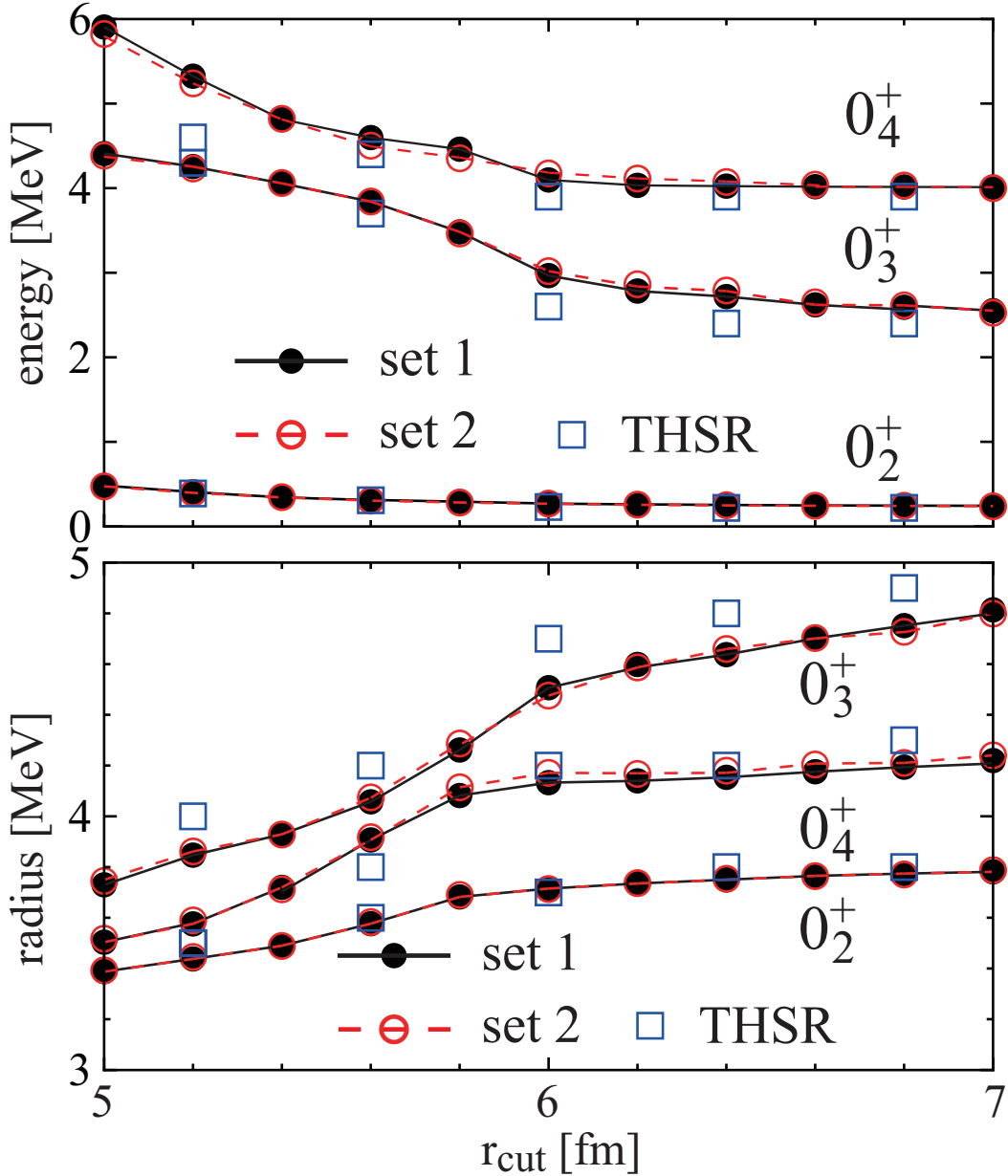}
  \caption{(Color online) energy (upper panel)  and radius (lower panel) of the excited $0^+$
  states obtained by the $r^2$-constraint. THSR results  are taken from  Ref. \cite{Funaki2015}} 
  \label{fig:er_cnst} 
 \end{center}
\end{figure}
Figure \ref{fig:er_cnst} shows the energies and radii of the excited $0^+$ states obtained by
the $r^2$-constraint. The energies of the $0^+_2$ and $0^+_4$ states are approximately constant in
the region of the $r_{cut} \geq 6$ fm, and the radii are very slowly increases as function of
$r_{cut}$. This implies that the most of the resonant wave functions in the interaction region is
already described by the basis wave functions with $r_{cut} < 6$ fm, and the choice of the 
$6 \leq r_{cut} \leq 7$ fm will give reasonable approximation for the $0^+_2$ and $0^+_4$
states. We also note the results for the  $0^+_2$ and $0^+_4$ look almost consistent with the THSR
results. On the other hand, we have not obtained the reasonable convergence for the $0^+_4$
state. In particular, the radius continues to increase as function of $r_{cut}$ not only in the
REM calculation but also in the THSR calculation, which implies the contamination non-resonant wave
functions. This requires more sophisticated method such as the complex scaling for more precise
discussion of this state \cite{Kurokawa2005, Kurokawa2007, Ohtsubo2013}.

\begin{figure}[h]
 \begin{center}
  \includegraphics[width=0.85\hsize]{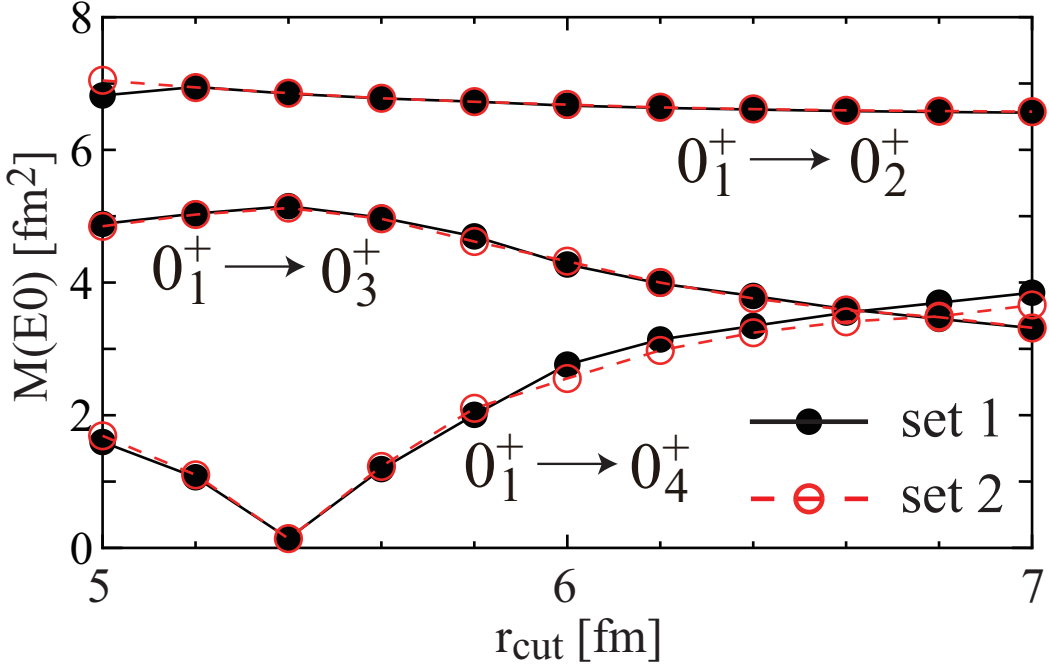}
  \caption{(Color online) Electric monopole transition matrix from the ground state to the excited
  $0^+$  states obtained by the $r^2$-constraint.}
  \label{fig:m0_cnst} 
 \end{center}
\end{figure}

Finally, Fig. \ref{fig:m0_cnst} shows the electric monopole transition matrix between the ground
and excited $0^+$ states. Again we see that the Hoyle state is quite stable, while the $0^+_3$ and
$0^+_4$ are dependent on $r_{cut}$. Since the monopole matrix element is very sensitive to the
outer side of the wave functions, this behavior also indicates the non-negligible contamination of
the continuum states with large radii.

\subsection{Excitation spectrum of $^{\bf 12}{\bf C}$}
\begin{figure*}[hbt]
 \begin{center}
  \includegraphics[width=0.9\hsize]{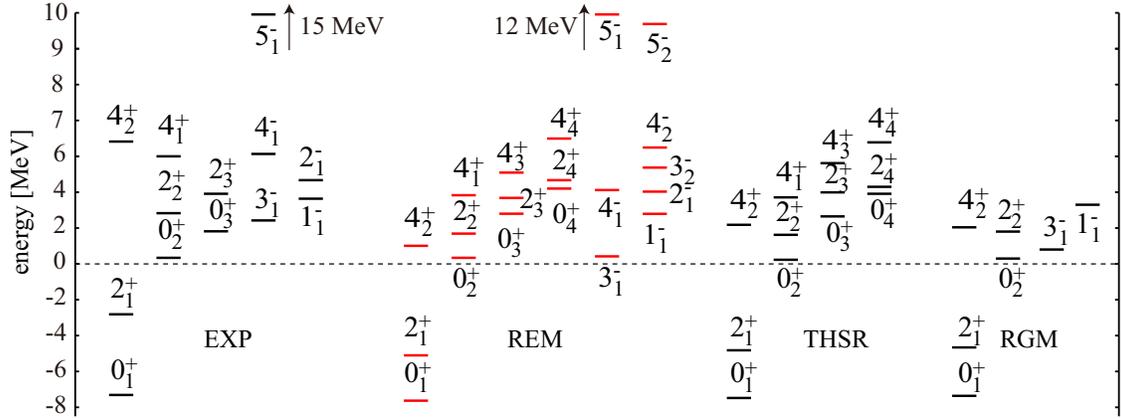}
  \caption{(Color online) Excitation energies measured from  the $3\alpha$ threshold. 
  Theoretical results by the REM, THSR \cite{Funaki2015,Funaki2016} and RGM
  \cite{Fujiwara1980,Kamimura1981} are compared with the experiments
  \cite{Ajzenberg-Selove1990,Freer2007,Kirsebom2010,Freer2011,Freer2012,Itoh2011,
  Zimmerman2011,Itoh2013a,Zimmerman2013,Marin-Lambarri2014}. The calculated $4^+_1$, $4^+_2$, 
  $5^-_1$ and $5^-_2$ states are labeled accordingly the observed counterparts.}
  \label{fig:spectrum}  
 \end{center}
\end{figure*}

Here, we discuss the excitation spectrum of $^{12}{\rm C}$ and make brief comments on the
structure of several states. Figure \ref{fig:spectrum} shows the excitation spectrum of 
$^{12}{\rm C}$ calculated by the REM with the ensemble set 1 and $r_{cut}=6.4$ fm together with
t hose by the RGM \cite{Fujiwara1980,Kamimura1981} and THSR \cite{Funaki2015,Funaki2016}. Their
energies and radii are also listed in Tab. \ref{tab:er}
Note that all of these calculations use the same Hamiltonian, and hence, they should be consistent
to each other and the deeper binding energy means a better wave function for the bound states. 
We see that all of the theoretical results are qualitatively consistent to each other. 
In particular, REM and THSR results are reasonably agree for all positive-parity states which
include the compact shell-model-like ground band and highly-excited cluster states. As for the
negative-parity states, REM and RGM reasonably agree for the $1^-_1$ and $3^-_1$ states, and REM
additionally produces the $2^-_1$, $3^-_2$ and $4^-_1$ states, which are also described by
AMD \cite{Kanada-Enyo1998,Kanada-Enyo2007a} and FMD \cite{Chernykh2007}. In short, REM can
describe all of the states reported by THSR and RGM reasonably. It must be emphasized that not
only the $0^+$ states but all of the state shown in Fig. \ref{fig:spectrum} were obtained from a
single ensemble set 1, which means that the EOM really effectively generates the basis wave
functions. 

\begin{table}[thb]
\caption{Calculated and observed energies measured from $3\alpha$ threshold in MeV and radii in
 fm. Experimental data is taken from Refs. \cite{Ajzenberg-Selove1990,Freer2007,Kirsebom2010,
 Freer2011,Freer2012,Itoh2011,Zimmerman2011,Itoh2013a,Zimmerman2013,Marin-Lambarri2014} }\label{tab:er} 
\begin{ruledtabular} 
\begin{tabular}{ccccccccc}
  & \multicolumn{2}{c}{REM} & \multicolumn{2}{c}{THSR \cite{Funaki2015,Funaki2016, Schuck2016}}
 & \multicolumn{2}{c}{RGM \cite{Fujiwara1980,Kamimura1981}} & \multicolumn{2}{c}{EXP}\\
 $J^\pi$& $E$ & $\sqrt{\braket{r^2}}$& $E$ & $\sqrt{\braket{r^2}}$& $E$ & $\sqrt{\braket{r^2}}$
 & $E$ & $\sqrt{\braket{r^2}}$\\\hline
 $0^+_1$& -7.6 & 2.4 & -7.5 & 2.4 & -7.4 & 2.4 & -7.3 & 2.4 \\
 $2^+_1$& -5.1 & 2.4 & -4.8 & 2.4 & -4.6 & 2.4 & -2.8 &  \\
 $4^+_2$&  1.0 & 2.3 &  2.2 & 2.3 &  2.0 & 2.3 &  6.8 &  \\
 $0^+_2$&  0.3 & 3.7 &  0.2 & 3.7 &  0.4 & 3.5 &  0.4 &  \\
 $2^+_2$&  1.7 & 3.9 &  1.6 & 3.9 &  2.1 & 4.0 &  2.8 &  \\
 $4^+_1$&  3.8 & 4.5 &  3.7 & 4.5 &      &     &  6.0 &  \\
 $0^+_3$&  2.8 & 4.6 &  2.7 & 4.7 &      &     &  1.8 &  \\
 $2^+_3$&  3.9 & 4.6 &  4.0 & 4.5 &      &     &  3.9 &  \\
 $4^+_3$&  5.4 & 4.8 &  5.6 & 4.7 &      &     &      &  \\
 $0^+_4$&  4.0 & 4.2 &  3.9 & 4.2 &      &     &      &  \\
 $2^+_4$&  4.6 & 3.7 &  4.3 & 4.1 &      &     &      &  \\
 $4^+_4$&  6.6 & 5.0 &  6.8 & 4.7 &      &     &      &  \\\hline
 $3^-_1$&  0.4 & 2.8 &      &     &  0.8 & 2.8 &  2.4 &  \\
 $4^-_1$&  4.1 & 2.9 &      &     &      &     &  6.1 &  \\
 $5^-_1$&  12  & 3.6 &      &     &      &     &  15  &  \\
 $1^-_1$&  2.8 & 4.3 &      &     &  3.4 & 3.4 &  3.6 &  \\
 $2^-_1$&  4.0 & 3.5 &      &     &      &     &  4.6 &  \\
 $3^-_2$&  5.4 & 4.5 &      &     &      &     &      &  \\
 $4^-_2$&  6.4 & 4.7 &      &     &      &     &      &  \\
 $5^-_2$&  9.3 & 4.5 &      &     &      &     &      &  
\end{tabular}
\end{ruledtabular}
\end{table}
Now, we discuss the details of the energies and radii listed in Tab. \ref{tab:er}. Firstly, for
many of the $2^+$ and $4^+$ states, we see that REM yields deeper binding energy than THSR and
RGM. This may be due to the limitation of the model space of the THSR and RGM
calculations. Namely, the THSR calculation assumes the axially symmetric intrinsic state and RGM
calculation limits the relative angular momentum between clusters up to 2, while REM has no such
assumptions. Secondly, THSR often yields smaller excitation energies and larger radii for highly
excited states such as $0^+_3$, $0^+_4$ and $2^+_4$ states. Although the variational principle
cannot be applied to these highly excited broad resonances, the difference may be attributed to
the difference in the long range part of the wave functions. THSR uses central Gaussians with
large size parameters, while REM uses localized Gaussians with relatively smaller size
parameters. Therefore, THSR should have better description for the long-range part of the  dilute
states. 

We also comment the difference between the models. In the THSR calculation, the positive-parity
condensate was assumed, hence the negative-parity is missing in the figure. However, if an
extended version of THSR is applied, we expect that it yields the negative-parity states
consistent with REM and RGM. In the RGM calculation, neither  $r^2$-constraint nor other
techniques to eliminate the non-resonant state were applied. As a result, it cannot describe many
highly excited states with large widths.  

Finally, we discuss the characteristics of the excited $0^+$ and $1^-$ states referring their
electric monopole and dipole transition strengths. The transition matrix are defined as,
\begin{align}
M(E0;0^+_m\rightarrow 0^+_n) &= \braket{0^+_n||\sum_{i=1}^Ar_i^{\prime 2}\frac{1+\tau_{zi}}{2}||0^+_m},\\
M(IS1;0^+_m\rightarrow 1^-_n) &= 
\braket{1^-_n||\sum_{i=1}^Ar_i^{\prime 3}Y_1(\hat{r}_i')||0^+_m},
\end{align}
where $\bm r_i'$ denotes the single-particle coordinate measured from the center-of-mass.
The results are summarized in Tab. \ref{tab:mono}.
\begin{table}[h]
\caption{Calculated and observed electric monopole and isoscalar dipole transition matrix in the
 units of $e\rm fm^2$ and $\rm fm^3$. }
 \label{tab:mono} 
\begin{ruledtabular}  
\begin{tabular}{ccccc}
 transition & REM & THSR \cite{Funaki2015,Funaki2016,Schuck2016} 
 & RGM  \cite{Fujiwara1980,Kamimura1981} & EXP \cite{Strehl1970}\\\hline
 $0^+_1\rightarrow 0^+_2$&  6.4  &  6.3 &  6.7 &  $5.4\pm 2$   \\
 $0^+_1\rightarrow 0^+_3$&  3.8  &  3.9 &      &        \\
 $0^+_1\rightarrow 0^+_4$&  3.3  &  3.5 &      &        \\
 $0^+_2\rightarrow 0^+_3$&  28   &  34  &      &        \\
 $0^+_2\rightarrow 0^+_4$&  0.7  &  0.5 &      &        \\
 $0^+_1\rightarrow 1^-_1$&  3.7  &      &      &        \\
 $0^+_2\rightarrow 1^-_1$&  45  &      &      &    
\end{tabular}
\end{ruledtabular}
\end{table}

\begin{figure}[h!]
 \begin{center}
  \includegraphics[width=0.8\hsize]{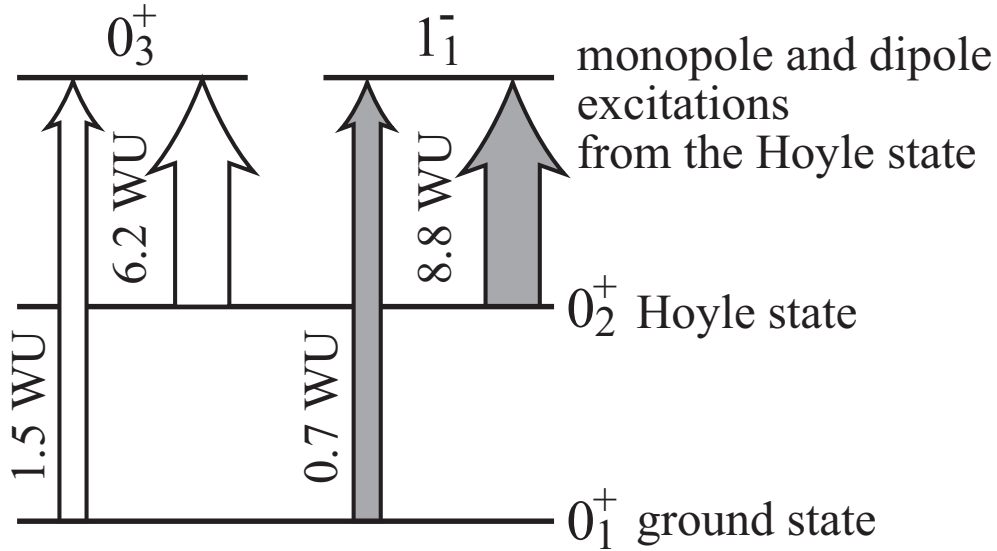}
  \caption{Excitation modes of the Hoyle state are schematically shown. Arrows show
  the monopole and dipole transitions.}   
  \label{fig:illust2}
 \end{center}
\end{figure}

Since the monopole transition operator is nothing but the radius operator, the matrix element
should be large for the dilute gas-like states \cite{Yamada2008}. Indeed, it is well known that
the Hoyle state has the enhanced monopole transition strength from the ground state because of its
dilute gas-like nature. The present calculation yields 6.4 $e\rm fm^2$ (1.5 WU) which is
consistent with the other cluster models, but slightly overestimates the observation.
The monopole transition from the ground state to the more dilute $0^+_3$ state is also large and
comparable with the Weisskopf unit (WU), but not as large as that of the Hoyle state. The reason
of the reduction is that the $0^+_3$  state is dominantly composed of the $4\hbar\omega$
configurations which cannot be excited by the monopole operator ($2\hbar\omega$ excitation).
However, it must be noted that the transition from the Hoyle state to the $0^+_3$ state is
greatly enhanced (6.2 WU). From this result and from the analysis of the wave function, it was
concluded that the $0^+_3$ state is a $2\hbar\omega$ excited state built on the Hoyle
state \cite{Funaki2015,Funaki2016}. In other words, it is the breathing mode of the Hoyle
state \cite{Zhou2016}. This relationship between the ground, Hoyle and $0^+_3$ states are
schematically  illustrated in Fig. \ref{fig:illust2}. On the contrary, the monopole transition between
the Hoyle state and the $0^+_4$ state is rather weak. This is due to the structural mismatch
between these states. In Ref. \cite{Funaki2016}, it was concluded that the $\alpha$ clusters are
linearly aligned in the $0^+_4$ state (linear-chain state), which naturally reduces the overlap
with the Hoyle state. 

A new finding in the present study is that not only the $0^+_3$ state but also the $1^-$ state may
be an excited state of the Hoyle state. As discussed in Ref. \cite{Chiba2016}, the $1^-$ states with
pronounced clustering should have the strong IS dipole transition strength from the ground
state. The present calculation yields  $M(IS1;0^+_1 \rightarrow 1^-_1)=3.7\ \rm fm^3$ which is
comparable with the Weisskopf unit (0.7 WU). A similar strength was also obtained by the AMD
calculation \cite{Kanada-Enyo2016a}.  Furthermore, it must be noted that the IS
dipole transition between  the Hoyle state and the $1^-_1$ state is extraordinary strong (8.8
WU). From this result, we are tempted to conclude that the $1^-_1$ state is the $1\hbar\omega$ (or
$3\hbar\omega$) excitation of the Hoyle state. Indeed the $1^-_1$ state has huge radius comparable
with the $0^+_3$ state. It is also interesting to note that the $1^-_1$ state is energetically
very close to the $0^+_3$ state. This conjecture is also illustrated in Fig. \ref{fig:illust2}. 

\begin{figure}[h!]
 \begin{center}
  \includegraphics[width=\hsize]{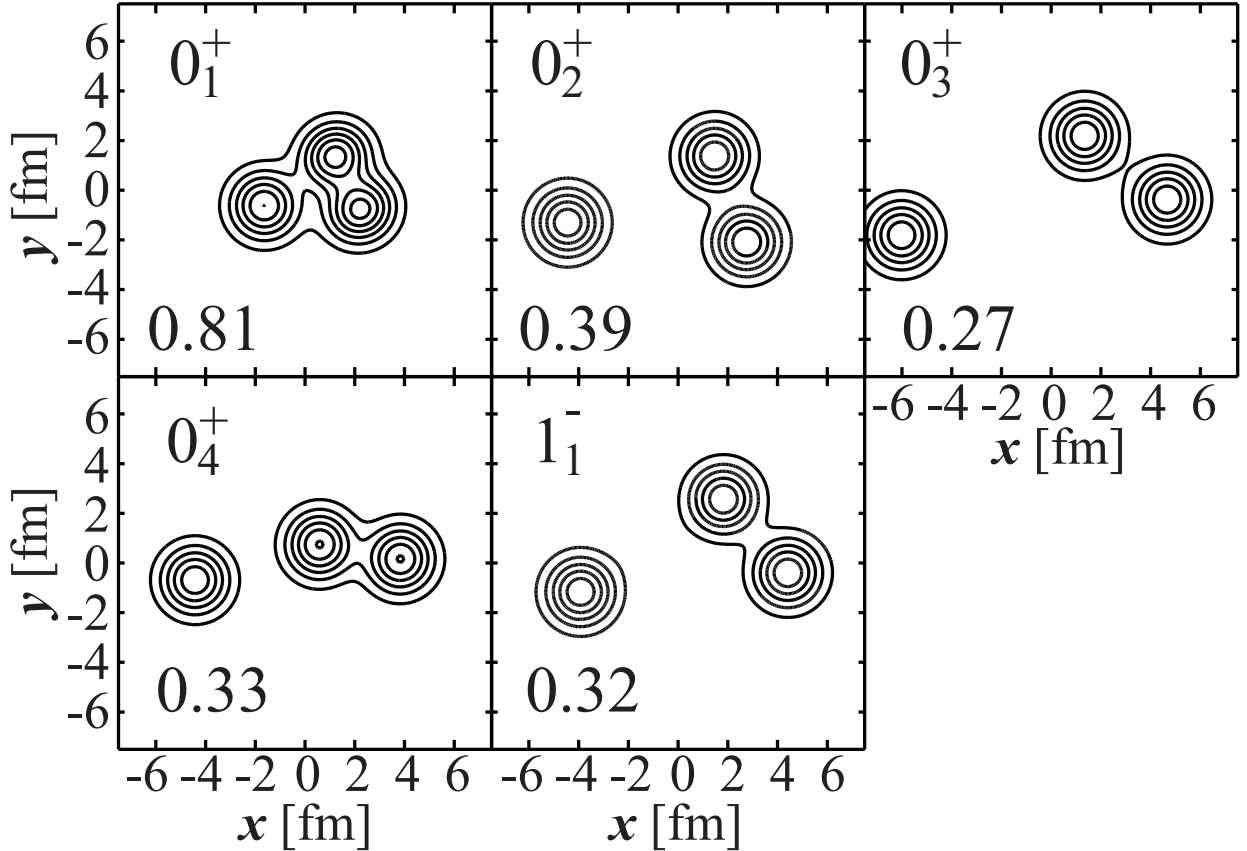}
  \caption{The most liklihood cluster configurations for the $0^+$ and $1^-$ states. Numbers in
  the panels show the maximum overlap defined by Eq. (\ref{eq:ovlp}).}  
  \label{fig:density2}
 \end{center}
\end{figure}

To discuss the similarity between the $0^+_3$ and $1^-_1$ states in a different way, 
we optimized a single Brink-Bloch wave function (optimized the position of 3$\alpha$
clusters) so that the overlap with the REM wave functions is maximized. Here the overlap is
defined as, 
\begin{align} 
 O = \frac{|\braket{\hat P^{J\pi}\Phi_{BB}|\Psi^{J\pi}}|^2}
 {\braket{\hat P^{J\pi}\Phi_{BB}|\hat P^{J\pi}\Phi_{BB}}},\label{eq:ovlp}
\end{align}
where $\Psi^{J\pi}$ denotes the REM wave function for the $0^+$ and $1^-$ states and $\Phi_{BB}$
is a Brink-Bloch wave function to be optimized. Thus-obtained optimized
Brink-Bloch wave functions shown in Fig. \ref{fig:density2} tell us the most likelihood cluster
configuration for each state. 
It can be seen that the ground state is represented by a Brink-Bloch wave function
with short inter-cluster distances whose overlap amounts to 0.81. This is due to the dominance of
the $(0s)^4(0p)^8$ configuration which can be described by a single Brink-Bloch wave function. On
the contrary, the most liklihood cluster configuration of the Hoyle state has large inter-cluster 
distance and it only explains about 39\%  of the Hoyle state. Indeed, not only the configuration
shown in Fig. \ref{fig:density2}, but many other configurations also have the overlaps of the same
magnitude. This indicates that the Hoyle state is not a single cluster configuration but a
superposition of many different configurations, which is consistent with the dilute nature of the
Hoyle state. The dilute character can be more strongly seen in the $0^+_3$ and $1^-_1$ states. The
likelihood configurations of these states have the inter-cluster distances larger than the Hoyle
state, and the overlaps are rather small. Thus, we expect that the $1^-_1$ state can be also
regarded as the family of the Hoyle state. We also comment that the $0^+_4$ state does not look a
perfect linear-chain configuration, but has a bent-armed configuration. The smallness of the
overlap may mean that this state is not stable against the bending motion. This interpretation is
consistent with the discussion made in Refs. \cite{Itagaki2001,Kanada-Enyo2007a,Funaki2016}. We
also note that this cluster configuration is more or less similar to the bent-armed cluster
configuration of the Hoyle state suggested by the Lattice calculation 
\cite{Epelbaum2012}.

\section{Summary}
In summary, we have developed a new theoretical model which  utilizes the classical EOM of the
Gaussian centroids to generate the  ergodic ensemble of the basis wave functions. The generated
basis wave functions are superposed to diagonalize the Hamiltonian. Thus, the method named REM can
be regarded a generator coordinate method which employs the real time $t$ as the generator
coordinate.  

As a benchmark of REM, we applied it to the 3$\alpha$ system ($^{12}{\rm C}$) and found that the
result is consistent with or even better than the other cluster models. It was shown that when the
propagation time is long enough the energies and radii of the ground and many excited states are
converged and independent of the initial condition. As a result, REM successfully described all of
the states reported by THSR and RGM. It must be emphasized that all the states are obtained from a
single ensemble of the basis wave function, which indicates that the EOM effectively generates the
basis wave functions. However, even if we apply the $r^2$-constraint, several excited states were
not converged well because of the contamination of the non-resonant wave function. Particular case
is the $0^+_3$ state which has broad width and is regarded as the breathing mode of the Hoyle
state. More accurate description of these states requires further development of the method.

Based on the isoscalar monopole and dipole transition strengths, the characteristics of the
excited $0^+$ and $1^-$ states are discussed. We confirmed that the properties of the Hoyle state
and the $0^+_3$ states are consistent with those discussed  in the preceding studies. They have 
dilute structure and the enhanced monopole transition strengths from the ground state. The huge
monopole transition strength between the Hoyle state and the $0^+_3$ state was also confirmed. In
addition to this, we found that the $1^-_1$ state has the analogous properties to the $0^+_3$
state. Namely, the $1^-_1$ state has dilute structure and the enhanced dipole transition strengths
from the ground state. It also has the extraordinary large IS dipole strength from the Hoyle
state. From these results, we conjecture that the $1^-_1$ state can be also regarded as an
excitation mode of the Hoyle state. Although this conjecture is based on only the transition
strengths and the overlaps,  more detailed quantitative discussion based on the reduced width
amplitudes, transition form factors and occupation probabilities will be made in our forthcoming
papers.  

\begin{acknowledgements}
The authors acknowledge that this work was initiated by the discussion with Dr. Kanada-En'yo and
Dr. Yabana. They also acknowledge the fruitful discussions with Dr. Zhou, Dr. Funaki, Dr. Horiuchi
and Dr. Kawabata. This work was supported by JSPS KAKENHI Grant No. 16K05339. 
\end{acknowledgements}

\bibliography{REM}
\end{document}